\documentclass[12pt]{article}%
\usepackage{amsmath,latexsym}
\usepackage{graphicx}
\usepackage{amsmath}
\usepackage{amsfonts}
\usepackage{amssymb}%\emph{\underline{}}
\begin{document}

\title{{A note on the stability of Morris-Thorne
    wormholes}}
   \author{
Peter K.F. Kuhfittig*\\  \footnote{kuhfitti@msoe.edu}
 \small Department of Mathematics, Milwaukee School of
Engineering,\\
\small Milwaukee, Wisconsin 53202-3109, USA}

\date{}
 \maketitle

\begin{abstract}\noindent
Two critical issues in the study of
Morris-Thorne wormholes concern the
stability of such structures, as well as
their compatibility with quantum field
theory.  This note discusses an important
subset characterized by zero tidal forces.
It is shown that such wormholes (1) are in
stable equilibrium using a criterion based
on the Tolman-Oppenheimer-Volkoff (TOV)
equation and (2) are not in direct conflict
with quantum field theory.

\end{abstract}

\section{Introduction}\label{S:introduction}

The Einstein field equations can be derived
by means of the Hilbert-Einstein action
\begin{equation}\label{E:HE}
  S_{\text{HE}}=\frac{1}{2\kappa}
     \int\sqrt{-g}\,R\,d^4x,
\end{equation}
where $R$ is the Ricci curvature scalar
and $\kappa=8\pi G$.  (For notational
convenience we will let $\kappa =1$.)
A modification of Einstein's theory
called $f(R)$ modified gravity
replaces $R$ by a nonlinear function
$f(R)$ in Eq. (\ref{E:HE}) to yield
\begin{equation}\label{E:f(R)}
   S_{f(R)}=\frac{1}{2\kappa}
      \int\sqrt{-g}\,f(R)\,d^4x.
\end{equation}
(For a review, see Refs. \cite{fL08,
 SF10, NO07}.)  Wormhole geometries in
 $f(R)$ modified gravitational theories
 are discussed in Ref. \cite{LO09}.

Here we need to recall that Morris and
Thorne \cite{MT88} proposed the following
static and spherically symmetric line
element for a wormhole spacetime:
\begin{equation}\label{E:line}
ds^{2}=-e^{2\Phi(r)}dt^{2}+\frac{dr^2}{1-b(r)/r}
+r^{2}(d\theta^{2}+\text{sin}^{2}\theta\,
d\phi^{2}).
\end{equation}
(We are using units in which $c=1$, in
addition to $\kappa =1$.)  The function
$b=b(r)$ is called the \emph{shape
function} and $\Phi=\Phi(r)$ is called the
\emph{redshift function}.  For the shape
function we must have $b(r_0)=r_0$, where
$r=r_0$ is the radius of the \emph{throat}
of the wormhole. Other requirements are
$b'(r_0)>0$ and $b'(r_0)<1$, called the
\emph{flare-out condition}, while $b(r)<r$
near the throat. In classical general
relativity, the flare-out condition can
only be satisfied by violating the null
energy condition (NEC):
\begin{equation}\label{E:NEC}
  T_{\mu\nu}k^{\mu}k^{\nu}\ge  0
\end{equation}
for all null vectors $k^{\mu}$, where
$T_{\mu\nu}$ is the stress-energy tensor.
In particular, for the outgoing null vector
$(1,1,0,0)$, the violation becomes
\begin{equation}\label{E:violation}
    T_{\mu\nu}k^{\mu}k^{\nu}=\rho +p_r<0.
\end{equation}
Here $T^t_{\phantom{tt}t}=-\rho$ is the energy
density, $T^r_{\phantom{rr}r}= p_r$ is the
radial pressure, and
$T^\theta_{\phantom{\theta\theta}\theta}=
T^\phi_{\phantom{\phi\phi}\phi}=p_t$ is
the lateral pressure.  In $f(R)$ modified
gravity, by contrast, meeting the flare-out
condition does not automatically result in
a violation of the NEC; so condition
(\ref{E:violation}) has to be checked
separately.

Regarding the redshift function, we normally
require that $\Phi(r)$ remain finite to
prevent the occurrence of an event horizon.
In this paper, we need to assume that
$\Phi(r)\equiv \text{constant}$, so that
$\Phi'(r)\equiv 0$.  Otherwise, according
to Lobo and Oliveira \cite{LO09}, the
analysis becomes intractable.  Fortunately,
the condition $\Phi'(r)\equiv 0$ is a
highly desirable feature for a traversable
wormhole since it implies that the tidal
forces are zero.

We can see from Eqs. (\ref{E:HE}) and
(\ref{E:f(R)}) that for the proper choice
of $f(R)$, the modified gravity theory can
be arbitrarily close to Einstein gravity.
For this choice, any Morris-Thorne
wormhole with zero tidal forces is found
to be stable by satisfying a well-known
equilibrium condition based on the
Tolman-Oppenheimer-Volkov equation.
%END OF SECTION

 \section{The solution}

As a first step, let us state the
gravitational field equations in the form
given in Ref. \cite{LO09}:
\begin{equation}\label{E:Lobo1}
   \rho(r)=F(r)\frac{b'(r)}{r^2},
\end{equation}
\begin{equation}\label{E:Lobo2}
   p_r(r)=-F(r)\frac{b(r)}{r^3}
   +F'(r)\frac{rb'(r)-b(r)}{2r^2}
   -F''(r)\left(1-\frac{b(r)}{r}\right),
\end{equation}
and
\begin{equation}\label{E:Lobo3}
   p_t(r)=-\frac{F'(r)}{r}\left(1-\frac{b(r)}{r}
   \right)+\frac{F(r)}{2r^3}[b(r)-rb'(r)],
\end{equation}
where $F=\frac{df}{dR}$.  The Ricci
curvature scalar is given by
\begin{equation}\label{E:R}
   R(r)=\frac{2b'(r)}{r^2}.
\end{equation}

In this paper, we are going to choose
\begin{equation}\label{E:modified}
   f(R)=aR^{1\pm\epsilon},\quad
       \epsilon\ll 1,
\end{equation}
where $a$ is a constant.  Given that
$\epsilon$ can be arbitrarily close
to zero, the resulting $f(R)$ modified
gravity can be arbitrarily close to
Einstein gravity.  Since $F=
\frac{df}{dR}$, we get from Eq.
(\ref{E:R})
\begin{equation}\label{E:F}
   F=a(1\pm\epsilon)R^{\pm\epsilon}=
   a(1\pm\epsilon)\left(\frac{2b'(r)}
   {r^2}\right)^{\pm\epsilon}.
\end{equation}
So from Eq.(\ref{E:Lobo1}),
\begin{equation*}
   b'=\frac{r^2\rho}{F}=\frac{r^2\rho}
   {a(1\pm\epsilon)(2/r^2)
   ^{\pm\varepsilon}
   [(b'(r)]^{\pm\epsilon}}
\end{equation*}
Solving for $b'$, we obtain
\begin{equation}
   b'(r)=\left(\frac{\rho}
   {2^{\pm\epsilon}a(1\pm\epsilon)}
   \right)^{\frac{1}{1\pm\epsilon}}r^2.
\end{equation}
If $a=1$ and $\epsilon =0$, we recover
Eq. (\ref{E:Lobo1}).  It also follows
from  Eq. (\ref{E:F}) that
\begin{equation}
   F=a(1\pm\epsilon)\left[2^{\pm\epsilon}
   \left(\frac{\rho}{2^{\pm\epsilon}
   a(1\pm\epsilon)}\right)^{\frac{\pm\epsilon}
   {1\pm\epsilon}}\right].
\end{equation}
Now observe that $F'$ has the form \
\[
     \frac{\pm\epsilon}{1\pm\epsilon}
     \rho^{\frac{\pm\epsilon}{1\pm\epsilon}-1}
     \times \,\,\text{a constant}.
\]
Since $\epsilon$ can be arbitrarily close
to zero, we conclude that
\begin{equation}
   F'\approx0 \quad \text{and} \quad
   F''\approx0.
\end{equation}

Our next task is to show that the NEC is
violated at the throat, i.e., $\rho(r_0)
+p_r(r_0)<0$.  (As already noted, meeting
the flare-out condition is not enough in
$f(R)$ modified gravity.)  To that end,
we first obtain
\begin{equation}\label{E:rho}
   \rho(r)=a(1\pm\epsilon)\left[2^{\pm\epsilon}
   \left(\frac{\rho}{2^{\pm\epsilon}a(1\pm\epsilon)}
   \right)^{\frac{\pm\epsilon}{1\pm\epsilon}}
   \right]\frac{b'(r)}{r^2},
\end{equation}
\begin{equation}\label{E:radial}
  p_r(r)=-a(1\pm\epsilon)\left[2^{\pm\epsilon}
  \left(\frac{\rho}{2^{\pm\epsilon}a(1\pm\epsilon)}
  \right)^{\frac{\pm\epsilon}{1\pm\epsilon}}
  \right]\frac{b(r)}{r^3},
\end{equation}
and
\begin{equation}\label{E:lateral}
   p_t(r)=a(1\pm\epsilon)\left[2^{\pm\epsilon}
   \left(\frac{\rho}{2^{\pm\epsilon}a(1\pm\epsilon)}
   \right)^{\frac{\pm\epsilon}{1\pm\epsilon}}\right]
   \frac{b(r)-rb'(r)}{2r^3}.
\end{equation}
We now get from $b(r_0)=r_0$,
\begin{equation}\label{E:WEC}
   \rho(r_0)+p_r(r_0)=a(1\pm\epsilon)\left[2^{\pm\epsilon}
   \left(\frac{\rho}{2^{\pm\epsilon}a(1\pm\epsilon)}
   \right)^{\frac{\pm\epsilon}{1\pm\epsilon}}\right]
   \left(\frac{b'(r_0)}{r^2_0}-\frac{1}{r^2_0}
   \right)<0
\end{equation}
since $b'(r_0)<1$.  The null energy condition
is thereby violated.
%END OF SECTION

\section{Stability analysis}\label{S:stability}

In this section we examine the stability of a
zero-tidal force wormhole by employing an
equilibrium condition obtained from the
Tolman-Oppenheimer-Volkov (TOV) equation
\cite{jP93, RKRI}
\begin{equation}\label{E:TOV}
   \frac{dp_r}{dr}+\Phi'(\rho+p_r)+
   \frac{2}{r}(p_r-p_t)=0.
\end{equation}
The equilibrium state of a structure is
determined from the three terms in this
equation, defined as follows: the
gravitational force
\begin{equation}
   F_g=-\Phi'(\rho+p_r),
\end{equation}
the hydrostatic force
\begin{equation}\label{E:hydrostatic}
   F_h=-\frac{dp_r}{dr},
\end{equation}
and the anisotropic force
\begin{equation}\label{E:anisotropic}
   F_a=\frac{2(p_t-p_r)}{r}
\end{equation}
due to the anisotropic pressure in a
Morris-Thorne wormhole.
Eq. (\ref{E:TOV}) then yields the
following equilibrium condition:
\begin{equation}
   F_g+F_h+F_a=0.
\end{equation}
Since $\Phi'\equiv 0$, the equilibrium
condition becomes
\begin{equation}
   F_h+F_a=0.
\end{equation}
So from Eqs.
(\ref{E:hydrostatic}) and (\ref{E:radial}),
we obtain
\begin{equation}
   F_h=-\frac{dp_r}{dr}=a(1\pm\epsilon)
   \left[2^{\pm\epsilon}\left(\frac{\rho}
   {2^{\pm\epsilon}a(1\pm\epsilon)}
   \right)^{\frac{\pm\epsilon}{1\pm\epsilon}}\right]
   \frac{r^3b'(r)-3r^2b(r)}{r^6},
\end{equation}
while Eqs. (\ref{E:radial}) and (\ref{E:lateral})
yield
\begin{equation}
   F_a=\frac{2}{r}(p_t-p_r)=\frac{2}{r}a(1\pm\epsilon)
   \left[2^{\pm\epsilon}\left(\frac{\rho}
   {2^{\pm\epsilon}a(1\pm\epsilon)}
   \right)^{\frac{\pm\epsilon}{1\pm\epsilon}}\right]
   \left(\frac{b(r)-rb'(r)}{2r^3}+\
      \frac{b(r)}{r^3}\right).
\end{equation}
The result is
\begin{multline}
   F_h+F_a=\\a(1\pm\epsilon)
   \left[2^{\pm\epsilon}\left(\frac{\rho}
   {2^{\pm\epsilon}a(1\pm\epsilon)}
   \right)^{\frac{\pm\epsilon}{1\pm\epsilon}}\right]
    \left(\frac{rb'(r)-3b(r)}{r^4}+
    \frac{b(r)-rb'(r)}{r^4}+\frac{2b(r)}{r^4}
    \right)=0.
\end{multline}
The equilibrium condition is satisfied, thereby
yielding a stable wormhole.  Since our modified
theory, based on Eq. (\ref{E:modified}), can be
arbitrarily close to Einstein's theory, the
conclusion carries over to Morris-Thorne
wormholes.

As a final comment, recall that the condition
$\Phi(r)\equiv \text{constant}$ was introduced
for purely technical reasons: without this
condition, the analysis becomes intractable.
So while the zero-tidal force assumption
proved to be a sufficient condition for
stability, it is not a necessary condition.
%END OF SECTION

\section{Compatibility with quantum field theory}
That quantum field theory may place severe
constraints on Morris-Thoene wormholes was
first shown by Ford and Roman \cite{FR96}.
The sought-after compatibility depends
on a quantum inequality in an inertial
Minkowski spacetime without boundary.  If
$u^{\mu}$ is the observer's four-velocity
and  $\langle T_{\mu\nu}u^{\mu}u^{\nu}\rangle$
is the expected value of the local energy
density in the observer's frame of reference,
then
\begin{equation}\label{E:QFT}
   \frac{\tau_0}{\pi}\int^{\infty}_{-\infty}
   \frac{\langle T_{\mu\nu}u^{\mu}u^{\nu}\rangle
   d\tau}{\tau^2+\tau_0^2}\ge
   -\frac{3}{32\pi^2\tau_0^4},
\end{equation}
where $\tau$ is the observer's proper time
and $\tau_0$ the duration of the sampling time.
(See Ref. \cite{FR96} for details.)  The
inequality can be applied in a curved
spacetime as long as $\tau_0$ is small
compared to the local proper radius of
curvature.  The desired estimates of the
local curvature are obtained from the
components of the Riemann curvature tensor
in classical general relativity, discussed
further in Ref. \cite{pK10}.

If we now let $F\equiv 1$ in Eqs.
(\ref{E:Lobo1})-(\ref{E:Lobo3}), we obtain
the Einstein field equations with
$\Phi'\equiv 0$.  As a result,  the
equilibrium condition in Sec.
\ref{S:stability} is automatically satisfied.
Unfortunately, according to Ref. \cite{pK10},
whenever $\Phi'\equiv 0$, the quantum
inequality is no longer satisfied and the
wormhole cannot exist on a macroscopic
scale.

Now the appeal to $f(R)$ modified gravity
becomes clear: as already noted, the
estimates of the local curvature needed
to apply Eq. (\ref{E:QFT}) come from
Einstein's theory, not from the modified
theory.  So the previous objections do
not apply.  More precisely, in the
equation $f(R)=aR^{1\pm\epsilon}$,
$\epsilon$ is always positive.  So
even if the modified theory is
arbitrarily close to Einstein's theory,
it remains an $f(R)$ theory, thereby
avoiding a direct conflict with
quantum field theory.
%END OF SECTION

\section{Conclusion}
While wormholes are a valid prediction of
Einstein's theory, their possible existence
faces additional challenges such as the
question of the stability of such structures,
as well as their compatibility with quantum
field theory.  This note addresses these
issues indirectly by invoking $f(R)$
modified gravity.  More precisely, the function
$f(R)$ in Eq. (\ref{E:modified}) leads to a
modification that can be arbitrarily close to
Einstein gravity.  It is subsequently shown that
a Morris-Thorne wormhole with zero tidal forces
is stable by satisfying an equilibrium condition
based on the Tolman-Oppenheimer-Volkov equation.
Since the modified theory can be arbitrarily
close to Einstein's theory, the conclusion
carries over to Morris-Thorne wormholes.  In
classical general relativity, however, the
zero-tidal force assumption is incompatible
with quantum field theory.  The $f(R)$ modified
theory based on Eq. (\ref{E:modified}) avoids
this problem since the constraints stemming
from Eq. (\ref{E:QFT}), being purely
classical, do not apply.

\end{document}